# Revival and splitting of a Gaussian beam in gradient index media

V. Arrizon,[1,*] F. Soto-Eguibar,[1] A. Zuñiga-Segundo,[2] and H.M. Moya-Cessa[1]

[1]Instituto Nacional de Astrofísica, Optica y Electrónica, Calle fiel Enrique Erro No. 1, Santa María Tonantzintla, Puebla, 72840 Mexico
[2]Departamento de Física, Escuela Superior de Física y Matemáticas, IPN,
*Corresponding author: arrizon@inaoep.mx

The short range revival of an arbitrary monochromatic optical field, which propagates in a quadratic GRIN rod, is a well-known effect that is established assuming the first-order approximation of the propagation operator. We discuss the revival and multiple splitting of an off-axis Gaussian beam propagating to relatively long distances in a quadratic GRIN medium. These effects are obtained assuming the second-order approximation of the propagation operator in this medium.

## 1.Introduction

A gradient index rod with quadratic index dependence in the radial coordinate (that we simply denote as GRIN), is usually employed in focusing and image formation [1]. An issue of theoretical and practical interest is that GRIN media can support invariant propagation modes, either in the paraxial [2] and the non-paraxial domains [3]. In a similar context, it has been established the appearance of self-images of periodic fields propagating in a GRIN rod [4, 5]. On the other hand, in a conceptual context, a GRIN medium have been employed in the formulation of the fractional Fourier Transform [6, 7].

An interesting effect is the revival of an arbitrary monochro-matic optical field, propagating in a GRIN medium, which has been derived assuming the first-order approximation of the propagation operator. For reasons to be exposed later, we will call this process as short period revival. We discuss another kind of propagation revival, which is shown by an off-axis Gaussian beam of appropriate parameters, during its propagation in a GRIN rod. The revival period in this process, is large in comparison to the period in the first-order case. Therefore the process is referred to as long period revival. Another interesting phenomenon that we also discuss in this context is the multiple splitting of the Gaussian beam, appearing at fractions of the revival length.

The long period revival and beam splitting in a GRIN rod have been previously established in [8]. It was done noting that the propagation operator in this device, approximated to second order, is formally equivalent to the evolution operator in a quantum Kerr medium, where the temporal revival and splitting had been established for coherent quantum states [9]. Such long period effects in the GRIN rod have been analyzed for one-dimensional (1-D) fields, and considering only the two beam splitting. Here we extend the discussion of such effects to the case of 2-D fields and consider multiple beam splitting. In section A we discuss the solution of the Helmholtz equation for a quadratic GRIN medium in terms of eigenfunctions, and establish the second-order approximation of the propagation operator. The long period revival and multiple beams splitting are analytically derived from this approximation. In section B we perform computational simulations to illustrate the splitting and revival of a Gaussian beam, propagating in a GRIN device, at specific conditions. The discussion of the results and concluding remarks are presented in section C.

### A.Theory

1.Propagation in a quadratic GRIN medium
Let us consider a transparent cylindrical rod (Fig. 1), whose refractive index is quadratic in the radial coordinate $r=(x^2+y^2)^{1/2}$, one of whose extreme faces is at the plane $z=0$. The Helmholtz equation for this medium can be expressed as

$$\frac{\partial^2 E}{\partial z^2} = -\left[\frac{\partial^2}{\partial x^2} + \frac{\partial^2}{\partial y^2} + k^2 n_0^2 [1 - g^2(x^2+y^2)]\right] E(x,y,z), \quad (1)$$

where $k$ is the wavenumber, $n_0$ is the axial refraction index, $g$ is the gradient index parameter, and $E(x,y,z)$ is the optical field.

Introducing the definitions $\kappa=kn_0$, $\eta=\kappa g$, and the number operators $\hat{n}_x(x)$ and $\hat{n}_y(x)$, such that $2\eta(\hat{n}_\xi + 1/2) = (-id/d\xi)^2 + \eta^2\xi^2$ ($\xi = x,y$), Eq. (1) can be rewritten as [10]

$$\frac{\partial^2 E}{\partial z^2} = -\left[\kappa^2 - 2\eta(\hat{n}_x + \hat{n}_y + 1)\right] E(x,y,z), \quad (2)$$

whose formal solution, in terms of the boundary condition $E(x,y,0)$, is

$$E(x,y,z) = \exp\left[iz\sqrt{\kappa^2 - 2\eta(\hat{n}_x + \hat{n}_y)}\right] E(x,y,0), \quad (3)$$

with $\tilde{\kappa}^2 = \kappa^2 - 2\eta$.

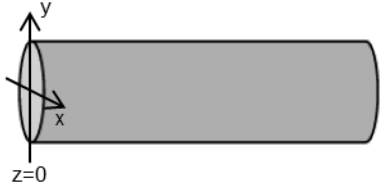

Fig. 1. Cylindrical GRIN rod.

To establish the second order approximation of the propagation operator in Eq. (3), we develop the square root as a Taylor series, and adopt only the first and second order terms, obtaining

$$E(x,y,z) = \exp\left\{-iz\left[\beta(\hat{n}_x + \hat{n}_y) + \gamma(\hat{n}_x + \hat{n}_y)^2\right]\right\}E(x,y,0). \quad (4)$$

where $\beta = \eta/\tilde{\kappa}$, and $\gamma = \eta^2/(2\tilde{\kappa}^3)$. The zero order term in the Taylor series, which is a constant phase factor, is omitted in Eq. (4).

For our further analysis we consider the eigenfunctions of the number operators $\hat{n}_x(x)$ and $\hat{n}_y(x)$ [11], given by

$$\varphi_n(\xi) = \left(\frac{\eta}{\pi}\right)^{1/4} \frac{1}{\sqrt{2^n n!}} \exp\left(-\frac{\eta}{2}\xi^2\right) H_n(\sqrt{\eta}\xi), \quad (5)$$
$$\xi = x, y, \quad n = 0,1,2...,$$

where $H_n$ are the Hermite polynomials. The boundary condition, expressed in terms of these eigenfunctions is

$$E(x,y,0) = \sum_{n=0}^{\infty}\sum_{m=0}^{\infty} c_{nm} \varphi_n(x) \varphi_m(y). \quad (6)$$

with coefficients

$$c_{nm} = \int_{-\infty}^{\infty}\int_{-\infty}^{\infty} E(x,y,0)\varphi_n(x)\varphi_m(y) dx\,dy. \quad (7)$$

Applying the propagation operator (Eq. (3)) to the boundary condition, we obtain the propagated field

$$E(x,y,z) = \sum_{n=0}^{\infty}\sum_{m=0}^{\infty} c_{nm} \varphi_n(x) \varphi_m(y) \exp\left[iz\sqrt{\tilde{\kappa}^2 - 2\eta(n+m)}\right]. \quad (8)$$

The second order approximation of this field, obtained by using the propagator in Eq. (4), is

$$E(x,y,z) = \sum_{n=0}^{\infty}\sum_{m=0}^{\infty} c_{nm} \varphi_n(x) \varphi_m(y) e^{-i\beta z(n+m)} e^{-i\gamma z(n+m)^2}. \quad (9)$$

In the first-order approximation the second complex exponential in Eq. (9) is removed and the well-known short range revival along z, with period $p_z = 2\pi/\beta$, is established.

2. Long range revival and multiple-split effects

The 1–D field for which the long period revival and splitting was formerly studied [8] is

$$\psi_\alpha(x) = \exp\left(-|\alpha|^2/2\right)\sum_{n=0}^{\infty} \frac{\alpha^n}{\sqrt{n!}} \varphi_n(x), \quad (10)$$

that has the structure of a quantum coherent state [12]. Using the Hermite polynomials exponential generating function $\exp(2xt - t^2) = \sum_{n=0}^{\infty} H_n(x)t^n/n!$ [11], valid for x and t complexes, it is shown that $\psi_\alpha(x)$ is the Gaussian function [13],

$$\psi_\alpha(x) = \sqrt[4]{\frac{\eta}{\pi}} \exp\left\{-\frac{\eta}{2}\left[x - \sqrt{\frac{2}{\eta}}\Re(\alpha)\right]^2\right\} \quad (11)$$
$$\exp\left\{i\sqrt{2\eta}\Im(\alpha)x - i\Re(\alpha)\Im(\alpha)\right\}$$

where $\Re(\alpha)$ and $\Im(\alpha)$ are the real and imaginary parts of $\alpha$, respectively. For the 2-D case to be discussed here we consider the boundary condition $\psi_\alpha(x,y) = \psi_\alpha(x)\psi_\alpha(y)$, given by

$$\psi_\alpha(x,y) = \exp\left(-|\alpha|^2\right)\sum_{n=0}^{\infty}\sum_{m=0}^{\infty} \frac{\alpha^{n+m}}{\sqrt{n!m!}} \varphi_n(x)\varphi_m(y). \quad (12)$$

According to Eq. (11), $\psi_\alpha(x,y)$ represents an off-axis Gaussian beam with equal displacements, proportional to $\Re(\alpha)$, along the x and y axes. The second order approximation for the propagated field [Eq. (9)], obtained for this boundary condition, is

$$E(x,y,z) = \exp\left(-|\alpha_N|^2\right)\sum_{n=0}^{\infty}\sum_{m=0}^{\infty} \frac{\alpha_N^{n+m}}{\sqrt{n!m!}} \varphi_n(x)\varphi_m(y)\exp[-i\gamma z(n+m)^2],$$
$$\alpha_N = \alpha\exp(-i\beta z). \quad (13)$$

A first interesting result occurs for the propagation distance $z = z_R = \pi/\gamma$. In this case the exponential within the sum in Eq. (13) becomes $(-1)^{n+m}$ and the propagated field can be expressed as

$$E(x,y,z_R) = \psi_{\alpha_R}(x,y), \quad (14)$$

where $\alpha_R = -\alpha\exp(-i\pi\beta/\gamma)$. Thus, the propagated field at the plane $z = z_R$, is a revival of the input field [Eq. (12)] with $\alpha$ replaced by $\alpha_R$.

Let us compare the revival period $z_R = \pi/\gamma$, with the period $p_z = 2\pi/\beta$, established in the first-order approximation. We first recall the definitions of parameters $\beta$ and $\gamma$, below Eq. (4), and the other related parameters, $\eta$, $\kappa$, and $\tilde{\kappa}$, to establish $p_z = 2\pi\tilde{\kappa}/(\kappa g)$ and $z_R = p_z \tilde{\kappa}^2/(\kappa g)$. Now, considering that the relation $g \ll \kappa$ is fulfilled (which is usual in practical cases), we can establish the approximations $\tilde{\kappa} \cong \kappa$, $p_z \cong 2\pi/g$, and $z_R \cong p_z(\kappa/g)$. As a consequence, we obtain $z_R \gg p_z$, which justifies calling $p_z$ and $z_R$ short and large revival periods, respectively.

Now we consider that $z = z_S = \pi/2\gamma$. In this case noting that the exponential within the sum in Eq. (13) is $(-1)^{n+m} i^{(n+m)^2}$, the propagated field is expressed

$$E(x,y,z_S) = e^{-|\alpha_S|^2}\sum_{n=0}^{\infty}\sum_{m=0}^{\infty} \frac{\alpha_S^{n+m}}{\sqrt{n!m!}} \varphi_n(x)\varphi_m(y) i^{(n+m)^2}, \quad (15)$$

where $\alpha_S = -\alpha\exp(-i\pi\beta/2\gamma)$. Noting that $i^{(n+m)^2}$ is either 1, for $n+m$ even, and $i$, for $n+m$ odd, Eq. (15) can be expressed as

$$E(x,y,z_S) = \frac{1}{2}e^{-|\alpha_S|^2}\sum_{n=0}^{\infty}\sum_{m=0}^{\infty}\frac{\alpha_S^{n+m}}{\sqrt{n!m!}}\varphi_n(x)\varphi_m(y)[1+(-1)^{n+m}]$$
$$+\frac{i}{2}e^{-|\alpha_S|^2}\sum_{n=0}^{\infty}\sum_{m=0}^{\infty}\frac{\alpha_S^{n+m}}{\sqrt{n!m!}}\varphi_n(x)\varphi_m(y)[1-(-1)^{n+m}]. \quad (16)$$

Performing some additional algebra in Eq. (16) we obtain

$$E(x,y,z_S) = \frac{1}{\sqrt{2}}\left[e^{i\pi/4}\psi_{\alpha_S}(x,y) + e^{-i\pi/4}\psi_{-\alpha_S}(x,y)\right]. \quad (17)$$

Therefore, the field propagated to the plane $z=z_S$, presents the 2-fold splitting of the input field $\psi_\alpha(x,y)$. Below Eq. (11) it is established that the a lateral shift of the 1-D Gaussian field $\psi_\alpha(x)$ is proportional to the real part of $\alpha$. Since this result also occurs for the 2-D Gaussian beam $\psi_\alpha(x,y)$, the two fields in Eq. (17) are symmetrically shifted with respect to the origin in the $(x,y)$ plane.

Extending the procedure employed in previous cases, we can expect the presence of multiple splitting of the Gaussian beam, considering a propagation distance $z=z_Q=\pi/Q\gamma$, with $Q>2$. Next we establish explicitly the 4-fold splitting, which is expected to occur at $z=z_4=\pi/4\gamma$. Considering this propagation distance in Eq. (13) we obtain

$$E(x,y,z_4) = \exp(-|\alpha_M|^2)\sum_{n=0}^{\infty}\sum_{m=0}^{\infty}\frac{\alpha_M^{n+m}}{\sqrt{n!m!}}\varphi_n(x)\varphi_m(y)\exp\left[-i\frac{\pi}{4}(n+m)^2\right], \quad (18)$$

where $\alpha_M = -\alpha\exp(-i\pi\beta/4\gamma)$. Next, noting that the complex expo-nential in Eq. (18) takes the values $i^{(n+m)}$, for $n+m$ even, and $\exp(-i\pi/4)$, for $n+m$ odd, the field can be expressed as

$$E(x,y,z_4) = \frac{1}{2}e^{-|\alpha_M|^2}\sum_{n=0}^{\infty}\sum_{m=0}^{\infty}\frac{(i\alpha_M)^{n+m}}{\sqrt{n!m!}}\varphi_n(x)\varphi_m(y)[1+(-1)^{n+m}]$$
$$+\frac{i}{2}e^{-|\alpha_M|^2}e^{-i\pi/4}\sum_{n=0}^{\infty}\sum_{m=0}^{\infty}\frac{\alpha_M^{n+m}}{\sqrt{n!m!}}\varphi_n(x)\varphi_m(y)[1-(-1)^{n+m}]. \quad (19)$$

Finally, employing the definition of the boundary condition in Eq. (12), the 4 implicit terms in Eq. (19) are expressed as

$$E(x,y,z_4) = \frac{1}{2}\psi_{i\alpha_M} + \frac{1}{2}\psi_{-i\alpha_M} + \frac{e^{-i\pi/4}}{2}\psi_{\alpha_M} + \frac{e^{-i\pi/4}}{2}\psi_{-\alpha_M}. \quad (20)$$

The field in Eq. (20) is formed by 4 replicas of the initial field $\psi_\alpha(x,y)$ with modified values for $\alpha$, given by $i\alpha_M, -i\alpha_M, \alpha_M$ and $-\alpha_M$. Recalling that the beam displacements are the real parts of the modified alphas and that $\alpha_M=-\alpha\exp(-i\pi\beta/4\gamma)$, the displacements of the beams in Eq. (20) are proportional to $\sin(\pi\beta/4\gamma), -\sin(\pi\beta/4\gamma), \cos(\pi\beta/4\gamma),$ and $-\cos(\pi\beta/4\gamma)$.

We point out that to predict the revival and split fields, it was necessary to take three terms in the Taylor expansion of the exact field propagation operator [Eq. (3)]. Such effects cannot be predicted by the first-order approximation alone and are difficult (if not impossible) to be obtained analytically from the exact propagation operator. Indeed, the presence of the second exponential in Eq. (9), quadratic in $(n+m)^2$, was explicitly required to obtain the mentioned effects.

**B. Numerical simulations**
The propagated field for different cases will be computed employing the paraxial, the second-order, and the exact approaches. In particular, we will show that the splitting and revival effects occur for restricted values of the parameters $\alpha$, $g$, and $\kappa$.

For an initial validation of our theoretical results obtained in previous section we consider the boundary condition in Eq. (12) with $\alpha=0$, which reduces to $\psi_\alpha(x,y) = \varphi_0(x)\varphi_0(y)$, the zero order eigenmode of the GRIN medium. The normalized intensity of this field is depicted in Fig. 2. The position units are normalized respect to the waist radius of the Gaussian beam that, according to Eq. (11), is $\omega_0=(2/\eta)^{1/2}$. It is obvious from Eqs. (8) and (9) that the intensity profile of the propagated field in this case, for any of the approaches, and arbitrary propagation distance $z$, is identical to that of the initial Gaussian beam (Fig. 2).

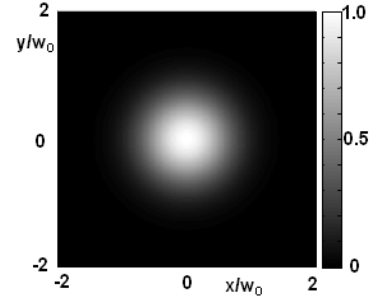

Fig. 2. Intensity of a Gaussian beam with null shift ($\alpha=0$).

In the following examples we will assume different values of the ratio $g/\kappa$. It will be shown that this ratio provides a measure of the relative weight of the different orders in the Taylor series of the propagation operator.

As second example, let us consider that the boundary condition is the coherent state [Eq. (12)] with a shift parameter $\alpha=2$, and that the GRIN medium parameters obey the relation $g/\kappa=10^{-4}$. The computed intensity of the field at the planes $z=n(p_z/4)$, $n=0...3$, employing the first-order approach, are displayed in Fig. 3. The field at $z=p_z$ (not shown) is identical to the one at $z=0$. As expected, during its propagation, the field oscillates around the optical axis, on a diagonal axis in the field of view, which is depicted in Fig. 3 (a). The position coordinates, shown in Fig. 3 (c) are identical for the four images. It is found that the fields computed with the second-order and exact approaches are, in this case, almost identical to those in Fig. 3. Fig. 4 shows the normalized field intensity profile on the diagonal oscillation axis, as a function of $z$, in the propagation range $[0, p_z]$. In this figure, and others, the position along the diagonal is denoted by $s$. The results in this second example are also consistent with already known facts in a GRIN medium [14].

As third example we consider the field around the plane $z=z_S$, where the two beam splitting is expected, maintaining the parameter $\alpha=2$ and the ratio $g/\kappa=10^{-4}$. The intensity of the field, computed with Eq. (7), at the planes $z=z_S+n(p_z/4)$, $n=-2, -1, 0,$ and 1, are displayed in Fig. 5. The intensity values, in the gray scale bars, are normalized respect to the peak intensity of the coherent mode (at plane $z=0$). It is noted in Fig. 5 (a,c) that in this case the original coherent mode [Fig. 3(a)] has been split into two identical modes, symmetrically arranged with respect to the optical axis. The two modes, that are oscillating around the optical axis, interfere at this axis at the planes $z=z_S-p_z/4$ and $z=z_S+p_z/4$ [Fig. 5 (b,d)]. The detailed oscillation of the field, depicted in Fig. 6, shows the field intensity profile on the diagonal oscillation axis, as a function of $z$, in the $z$ range $[z_S-p_z/2,\ z_S+p_z/2]$. The values of $z/p_z$, in figure 6, correspond to the gradient index $g=10^4\ m^{-1}$. Quite similar results are obtained in this case using the second–order approach. However, the use of the first-order approach in this case gives results similar to those in Fig. 3 that obviously are not correct.

Now let us consider the revival field. The field intensity profile on the diagonal axis, as a function of $z$, in the propagation range $[z_R-p_z/2, z_R+p_z/2]$, computed with the exact approach, is displayed in Fig. 7. In this figure, the values of $z/p_z$, also correspond to $g=10^4\ m^{-1}$. It is noted that this intensity profile shows small but visible differences respect to the field obtained near the plane $z=0$ [Fig. 4]. In particular, the intensity at the revival plane $z=z_R$, is quite similar to that at $z=0$. Thus, the exact computation in this case also corresponds to the prediction of the second-order approach.

We point out that the appearance of the split and revival effects depends on the ratio $g/\kappa$. In the previous numerical evaluations of these effects, the chosen ratio $g/\kappa=10^{-4}$ allows the appearance of the effects, for the shift parameter $\alpha=2$. If we assume, for instance, $g/\kappa=10^{-3}$ and $\alpha=2$, the field intensities at the planes $z=z_S$ and $z=z_R$, computed with the exact approach, are those displayed in Fig. 8 (a,b). Although the split and revival effects still occur in this case, the replicas of the initial coherent state, in the propagated fields appear distorted. If we assume $g/\kappa=10^{-2}$, and $\alpha=2$, the field intensities at the splitting and revival planes, computed with the exact approach, and displayed in Fig. 8 (c, d), indicate that the predicted effects are no longer verified. We point out that the computation with the second–order approach still provides the previously obtained results [figures 3 (a) and 5 (a)]. However, in the present case the prediction of the second-order approach has not been confirmed by the exact computation.

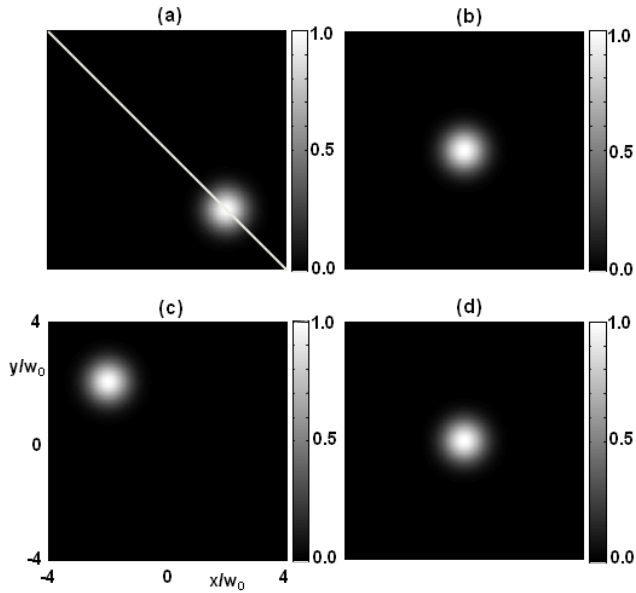

Fig. 3. Transverse oscillation of the input Gaussian field with shift $\alpha=2$, and parameters ratio $g/\kappa=10^{-4}$. The depicted intensities correspond to the propagation distance (a) $z=0$, (b) $z=p_z/4$, (c) $z=p_z/2$, and (d) $z=3p_z/4$.

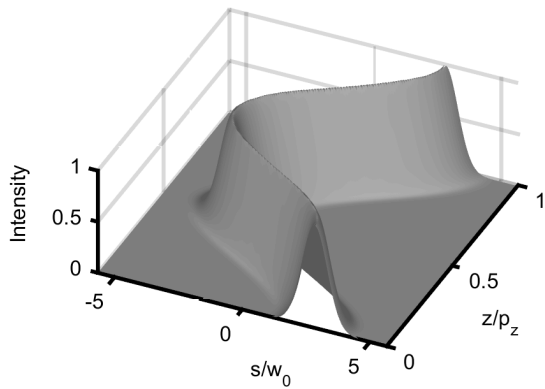

Fig. 4. Normalized field intensity on the diagonal oscillation axis (with positional coordinate $s$) versus $z$, in the range $[0, p_z]$, for the input Gaussian field shift $\alpha=2$, and parameters ratio $g/\kappa=10^{-4}$.

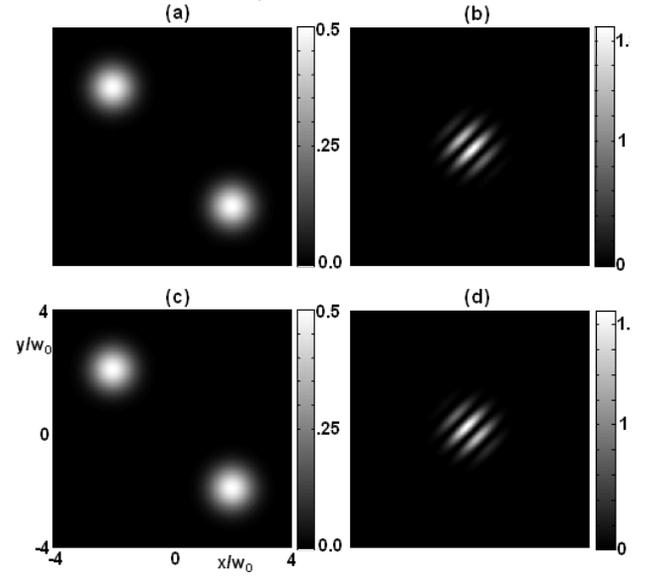

Fig. 5. Two beam splitting of the input Gaussian beam with parameters $\alpha=2$, and $g/\kappa=10^{-4}$. The depicted intensities correspond to the propagation distance (a) $z_S-p_z/2$, (b) $z_S-p_z/4$, (c) $z_S$, and (d) $z_S+p_z/4$.

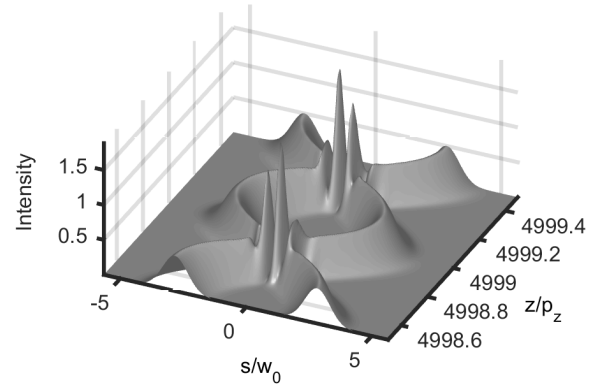

Fig. 6. Normalized field intensity on the diagonal oscillation axis versus $z$, in the range $[z_S-p_z, z_S+p_z]$, for the input Gaussian field shift $\alpha=2$, and parameters ratio $g/\kappa=10^{-4}$.

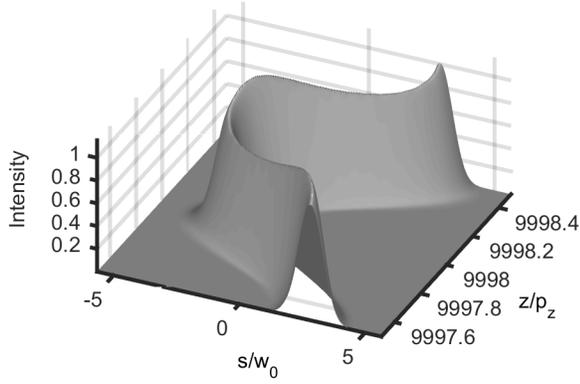

Fig. 7. Normalized field intensity on the diagonal oscillation axis versus $z$, in the range $[z_R-p_z, z_R+p_z]$, for the input Gaussian field shift $\alpha=2$, and parameters ratio $g/\kappa=10^{-4}$..

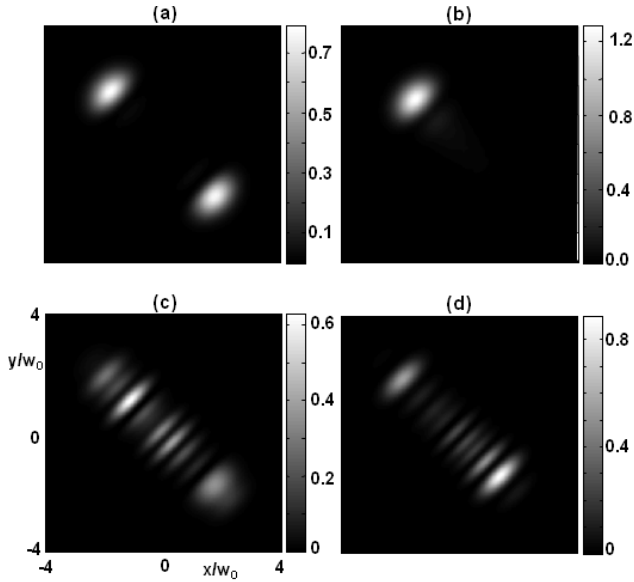

Fig. 8. Two beam splitting (a) and revival (b) of the input Gaussian beam with parameters $\alpha=2$, and $g/\kappa=10^{-3}$. These effects disappear (c, d) when the ratio $g/\kappa$ is increased to $10^{-2}$.

There is a dependence between the ratio $g/\kappa$, required to obtain the split and revival effects, and the value of the shift parameter $\alpha$. Increased values of $\alpha$ require smaller ratios $g/\kappa$. Let us consider, as an example, the splitting fields for a shift parameter $\alpha=4$, and two different values of $g/\kappa$: $10^{-4}$ and $10^{-5}$. The intensities of the computed fields at the splitting distance ($z=z_S$), displayed in Fig. 9, show that the split fields appear distorted for $g/\kappa=10^{-4}$ and take the appropriate shape for $g/\kappa=10^{-5}$.

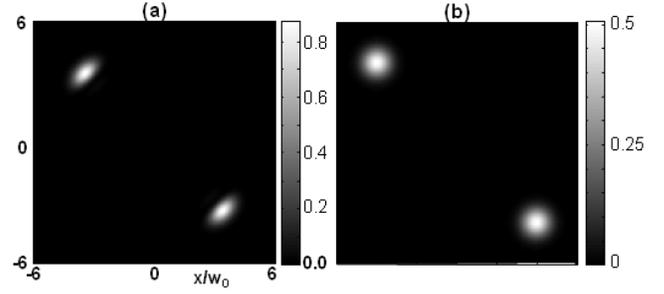

Fig. 9. Two beam splitting of the input Gaussian beam with a shift parameter $\alpha=4$ and ratio $g/\kappa$ with values (a) $10^{-4}$ and (b) $10^{-5}$.

We also computed the field obtained at the distance $z=\pi/4\gamma$, for which the 4-fold splitting [Eq. (20)] is expected. The intensity field at this plane, computed for the parameters $\alpha=2$ and $g/\kappa=10^{-4}$, and displayed in Fig. 10 (a), seems to be formed by 4 shifted replicas of the initial coherent state. Two of these replicas are interfering at the optical axis and the other two appear at the extremes of the image. The 4-fold splitting is more visible if we take a new shift parameter $\alpha=4$, assuming $g/\kappa=10^{-5}$. In this case, the field intensity at the plane $z=\pi/4\gamma-(p_z/16)$ [Fig. 10 (b)] clearly shows the four shifted replicas of the input Gaussian beam. The shift $p_z/16$ (introduced in $z$) is required to obtain the replicas of the Gaussian beam uniformly distributed in the field of view. Thus, the fields in Fig. 10, computed with the exact approach, confirms the 4 beam splitting predicted by the second-order approach.

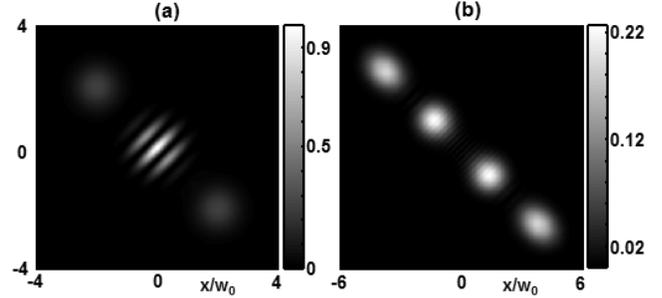

Fig. 10. Four beam splitting for the combinations of parameters (a) ($\alpha=2, g/\kappa=10^{-4}$), and (b) ($\alpha=4, g/\kappa=10^{-5}$). The propagation distances are $z=\pi/4\gamma$ and $z=\pi/4\gamma-(p_z/16)$, respectively.

### C. Discussion and final remarks

Let us briefly analyze the conditions for validity of the second order approximation of the propagated field [Eq. (13)]. In order to establish such conditions it is convenient to express the propagated field employing the propagation operator approximated to third order. In this case the propagated field is given by

$$E(x,y,z) = \exp\left(-|\alpha|^2\right)\sum_{n=0}^{\infty}\sum_{m=0}^{\infty}\frac{\alpha^{n+m}}{\sqrt{n!m!}}\varphi_n(x)\varphi_m(y) \quad (21)$$
$$\exp\left\{-iz[\beta(n+m)+\gamma(n+m)^2+\xi(n+m)^3]\right\},$$

where $\xi = \eta^3/(2\tilde{\kappa}^5)$. We note that the second-order approximation [Eq. (13)] is valid when the third term, $\xi(n+m)^3$, in the exponential of the propagation factor in Eq. (21), is negligible respect to the second term, $\gamma(n+m)^2$. In other words we require that the quotient of these terms, $R=(n+m)\xi/\gamma$, be much smaller than unity. Now, recalling the expressions $\xi=\eta^3/(2\tilde{\kappa}^5)$ and $\gamma=\eta^2/(2\tilde{\kappa}^3)$, and assuming $g/\kappa<<1$ (as is the case in the considered examples) it is found that $\xi/\gamma<<1$. On the other hand, it is noted that the first factor in $R$, $(n+m)$, increases for large order modes (in Eq. (21)). In addition, the necessity of considering high order modes increases for high values of $\alpha$. This fact explains why the second factor in $R$, ($\approx g/\kappa$), must be reduced while $\alpha$ increases. In the examples considered in section B we have shown numerically that the splitting and revival effects occur appropriately in a GRIN medium adopting $g/\kappa<10^{-4}$ for $\alpha=2$ and $g/\kappa<10^{-5}$ for $\alpha=4$.

Let us evaluate the waist radius of the input Gaussian beam, which, as stated in section B, is $\omega_0=(2/\eta)^{1/2}$. Assuming, e. g. the axial refractive index $n_0=1.5$ and the parameter quotient $g/\kappa$ equal to $10^{-4}$ and $10^{-5}$, employed in the numerical simulations, the waist radius of the coherent state adopt the values $\omega_0\approx15\lambda$ and $\omega_0\approx47\lambda$ respectively. These relatively small radii of the input beams are still reasonable from the physical point of view.

Now we compute the propagation distances at which the revival and splitting effects can be observed. First we note that considering the quotient $g/\kappa=10^{-4}$, the relations $z_R=2z_S=p_z(\kappa/g)$ are approximately valid, and we obtain that the splitting and revival effects occur at propagation distances $z_S=5000p_z$ and $z_R=10000p_z$, respectively, where $p_z=2\pi/\beta$ is the short period revival. Such large distances appear at the simulations illustrated in figures 6 and 7. On the other hand, considering $g/\kappa<<1$, the period $p_z$ is approximated to $2\pi/g$. In addition, according to published data [15–17], the values of the gradient parameter $g$, for conventional GRIN devices operating in the visible domain, are in the range from $10^2\,m^{-1}$ to $10^4\,m^{-1}$. For the extreme gradient index $g=10^4\,m^{-1}$, and the assumed quotient $g/\kappa=10^{-4}$, we obtain the short revival period $p_z=2\pi\times10^{-4}\,m$, while the large 2-fold split and revival distances are $z_S\approx3.14\,m$ and $z_R\approx6.28\,m$ respectively. Such propagation distances are too large for conventional GRIN devices. If $n_0=1.5$, the required wavelength, for the considered parameters, is $\lambda\approx.094\,\mu m$, which belongs to the UV range. Since assuming the condition $g/\kappa<<1$, we established $z_R\approx(\kappa/g)p_z$, the distance $z_R$ can be reduced by taking smaller values for $g/\kappa$. E. g. maintaining $g=10^4\,m^{-1}$ and taking $g/\kappa=10^{-2}$ one obtains $z_S\approx3.14\,cm$ and $z_R\approx6.28\,cm$. The required wavelength in this last case is $\lambda\approx9.4\,\mu m$, which belongs to the medium IR range.

In summary, we have discussed the large period revival and multiple beam splitting of an appropriately shaped and shifted Gaussian beam, propagated in a GRIN medium. Such effects are predicted by means of a propagation operator approximated to second order. The propagation distances required to observe the large period revival and split effects are large in comparison to lengths of available GRIN rods. Therefore, the experimental implementation and application of such effects is a challenging task to be realized.